\documentclass[usenatbib,usegraphicx,useAMS]{mn2e}
\usepackage{amsmath}
\usepackage{amssymb}

\newcommand{\Ms}{M$_\odot$ }

\def\eg{{e.g.}}
\def\ie{{i.e.}}

\def\etc{{etc.}}
 
\title{A multi-dimensional, adiabatic hydrodynamics code for
studying tidal excitation}
\author[Avery E.~Broderick and Yasser Rathore]{Avery E.~Broderick$^1$\thanks{E-mail: abroderick@cfa.harvard.edu}
and Yasser Rathore$^2$\thanks{E-mail: yasser@caltech.edu} \vspace{0.1cm} \\
$^1$ Institute for Theory and Computation, Harvard-Smithsonian
Center for Astrophysics, Cambridge, MA 02138, USA \\
$^2$ Theoretical Astrophysics, Caltech 130-33, Pasadena, CA 91125, USA}

\begin{document}
\maketitle

\begin{abstract}
We have developed a parallel, simple, and fast hydrodynamics
code for multi-dimensional, self-gravitating, adiabatic flows.  
Our primary motivation is the study of the non-linear evolution of
white dwarf oscillations excited via tidal resonances, typically over
hundreds of stellar dynamical times.  Consequently, we
require long term stability, low diffusivity, and high algorithmic
efficiency.  An Eulerian finite-difference scheme on a regular Cartesian grid
fulfills these requirements.  It provides uniform resolution throughout the flow,
as well as simplifying the computation of the self-gravitational potential, which
is done via spectral methods.  In this paper, we describe the numerical scheme and
present the results of some diagnostic problems.  We also demonstrate the stability
of a cold white dwarf in three dimensions over hundreds of dynamical times.
Finally, we compare the results of the numerical scheme to the linear theory of
adiabatic oscillations, finding numerical quality factors on the order of 6000, and
excellent agreement with the oscillation frequency obtained by the linear analysis.
\end{abstract}

\begin{keywords}
hydrodynamics -- methods: numerical -- white dwarfs -- binaries: close
\end{keywords}

\section{Introduction} \label{Intro}
White dwarfs (WDs) are known to be common endpoints of stellar evolution.  A
significant amount of evidence suggests that both stellar mass black holes
($<10^2$ \Ms) and neutron stars are also relatively common.  More recently,
both theoretical \citep[see, \eg,][]{Mada-Rees:01} and observational
\citep[see, \eg,][]{Colb-Ptak:02,Gers_etal:02} studies have implied
the existence of
intermediate mass black holes ($10^2$--$10^5$\Ms).  As a result, it
appears inevitable that white dwarf--compact object binaries will form.
This may be especially likely within cluster environments.

After formation, the subsequent evolution of a white dwarf--compact object binary
will typically be driven by gravitational radiation.  As the system passes through
resonances between the normal mode frequencies of the white dwarf and
harmonics of the orbital frequency, it is possible to resonantly excite
oscillations on the white dwarf.  Even small amounts of energy transfer
may have a non-negligible impact upon the orbit, possibly with consequences
for gravitational wave detections of such systems (\eg, by LISA).
Large energy transfers may result in heating and, possibly, the detonation
of the white dwarf, leading to an exotic type I supernova and, perhaps, a subsequent
$\gamma$-ray burst.  In order to assess the magnitude and likelihood of such scenarios,
it is necessary to understand the mode excitation process in detail.  For the linear
regime, this has been done \citep{Rath-Brod-Blan:03, Rath-Blan-Brod:04}, and it was
found that, depending upon the initial conditions, it is possible to excite modes with
large enough amplitudes that the validity of the linear theory becomes questionable.
Therefore, it is necessary to investigate the mode evolution in the non-linear regime.
This is most directly done via numerical hydrodynamics simulations.

A number of hydrodynamics codes which may be used for this purpose
currently exist.  Two such codes, ZEUS \citep{Ston-Norm:92} and Flash
\citep{Flash:00} have been developed to be generic hydrodynamic
engines.  Such codes provide access to a sophisticated suite of hydrodynamic
simulation tools.  However, they also have the disadvantage of being
complicated to use and, perhaps, suboptimal for our specific problem.  In
addition, to a good approximation, the white dwarf oscillations are
adiabatic, and hence detailed treatment of shocks and entropy
generation are unnecessary.

\citet{Motl-Tohl-Fran:02} have developed an adiabatic hydrodynamics
code, primarily for studying binary mass transfer.  However, the choice
of a cylindrical grid, while useful for the mass transfer application,
is problematic for the case of a pulsating white dwarf, where it is important
to maintain uniform resolution throughout the star.  Furthermore, a cylindrical
coordinate system complicates the numerical advection scheme.
These difficulties are avoided with a Cartesian grid, an additional advantage
of which is that the Poisson equation can be solved easily and efficiently
via spectral methods.

We present a simple hydrodynamics code with some diagnostics and an
example application.  This is done in seven sections with
\S\ref{GHE} reviewing the hydrodynamic equations, \S\ref{DS} outlining
the differencing scheme used, \S\ref{StPE} describing the method used
to solve the Poisson equation, \S\ref{TP} presenting some tests of the
code, \S\ref{AtaPWD} applying the code to an oscillating white dwarf,
and \S\ref{C} containing concluding remarks.

\section{Governing Hydrodynamic Equations} \label{GHE}
There is considerable freedom in the choice of macroscopic quantities
used to describe fluid flows.  Our choice was primarily dictated by the
numerical convenience of the sourced advective form of the hydrodynamic
equations.  In addition, since we are restricting ourselves to
adiabatic flows, it is convenient to use the entropy rather than the
energy as a thermodynamic variable. We therefore chose the following
five quantities to describe the fluid flow: mass density ($\rho$),
entropy density ($s$), and momentum density (${\bf J}$).

The equations for $\rho$ and $s$ have a purely advective form,
\begin{align}
\frac{\partial \rho}{\partial t} + \bmath{\nabla} \cdot {\bf v} \rho &= 0 
\label{mass_cons} \\
\frac{\partial s}{\partial t} + \bmath{\nabla} \cdot {\bf v} s &= 0 \,,
\label{entropy_cons}
\end{align}
which correspond to the conservation of mass and entropy.\footnote{Note
that $s$ is the entropy per unit volume and not the specific
entropy. Hence, in our notation, the adiabatic condition is
\begin{equation*}
\frac{d}{d t}\left(\frac{s}{\rho}\right) = 0 \,,
\end{equation*}
where $d/dt$ is the convective derivative.} The equation for ${\bf J}$
can be written in a sourced advective form,
\begin{equation}
\frac{\partial {\bf J}}{\partial t} + \bmath{\nabla} \cdot {\bf v} \, {\bf J}
=
-\bmath{\nabla} P - \rho \bmath{\nabla} \Phi + {\bf f}
\,,
\label{momentum_cons}
\end{equation}
where the pressure ($P$) is given by an equation of state,
\begin{equation}
P = P(\rho,s) \,,
\label{eos}
\end{equation}
the self-gravitational potential ($\Phi$) is determined by the Poisson
equation,
\begin{equation}
\nabla^2 \Phi = 4 \pi G \rho \,,
\label{poisson_eq}
\end{equation}
and ${\bf f}$ is any additional external force per unit volume acting
on the fluid (\eg, an external gravitational field and/or Coriolis forces).

\section{Differencing Scheme} \label{DS}
In one dimension, the use of
a staggered mesh avoids the interpolation of the flow velocities to the cell
boundaries. With a zone-centred grid, the velocities would have to be
interpolated, which would complicate the advection step in the momentum
conservation equation (\ref{momentum_cons}). However, in multiple dimensions,
the interpolation of vector quantities (\eg, the momentum density) cannot be avoided
by the use of a staggered mesh.  Therefore, we use the conceptually simpler
zone-centred grid.

\begin{figure}
\begin{center}
\includegraphics[width=\columnwidth]{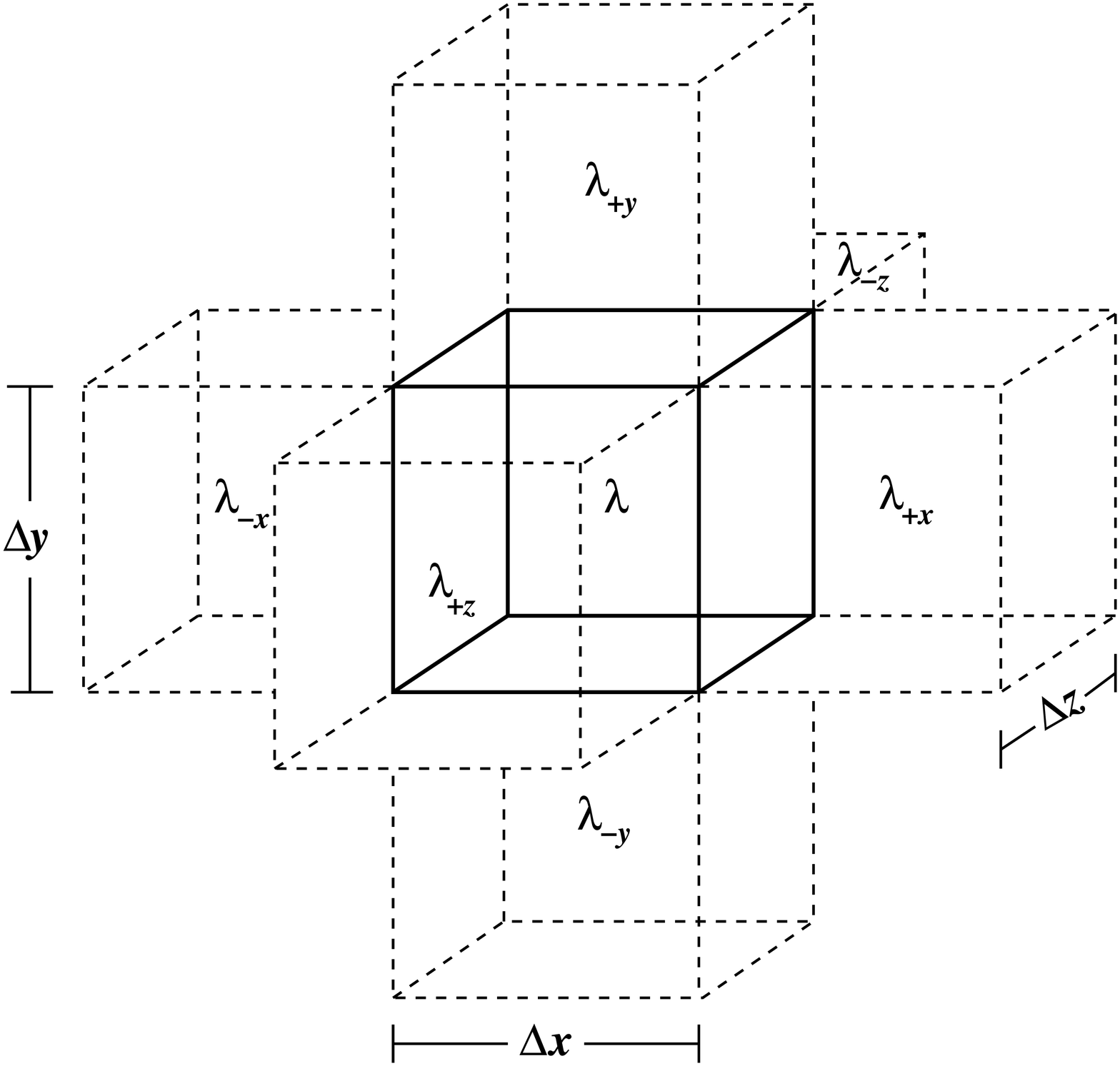}
\end{center}
\caption{
The geometry of a zone-centred, uniform Cartesian grid is shown.
Here, $\lambda$ can be any of the five evolved quantities
($\rho$, $s$, and ${\bf J}$) or the gravitational potential ($\Phi$).
}
\label{diagram}
\end{figure}

Casting the hydrodynamic equations in a sourced advective form
allows the explicit conservation of mass, entropy, and momentum
(insofar as the source terms allow).

\subsection{Advection}
The advection steps in equations (\ref{mass_cons}--\ref{momentum_cons}) may be
integrated to yield finite-difference equations for a given cell
\begin{equation}
\Delta \lambda = \frac{\Delta t}{\Delta V} \sum_{i=x,y,z} \left(
\Lambda_{-i} 
-\Lambda_{+i}
\right) \Delta S_i \,,
\end{equation}
where $\Delta \lambda$ is the change in the quantity $\lambda$ due to
fluid advection, $\Delta t$ is the time step, $\Delta V$ is the cell volume,
$\Lambda_{\pm i}$ are the fluxes of the quantity $\lambda$ at the $\pm i$th
boundary of the cell, and $\Delta S_i$ is the area of the cell surface normal
to the $i$th direction.

In general some interpolation is required to
determine the values of the fluxes at the boundaries of the cell.
We break the interpolation of the fluxes into an interpolation over the
fluid velocity and an interpolation over the advected quantities,
\begin{equation}
\Lambda_{\pm i}
=
\lambda^*_{\pm i} \overline{v}_{\pm i} \,,
\end{equation}
where $\overline{v}_{\pm i}$ is the interpolated component of the velocity
normal to the $\pm i$th cell face at the cell boundary, 
and $\lambda^*$ is the interpolated value of the
advected quantity.  The $\overline{v}_{\pm i}$ are defined by
\begin{equation}
\overline{v}_{\pm i} = \frac12 ( v^i + v^i_{\pm i} ) \,,
\end{equation}
where $v^i$ and $v^i_{\pm i}$ are the values of the fluid velocity in the
$i$th direction at the centre of the current cell, and the centres of the
neighbouring cells in the $\pm i$th directions, respectively.

A numerical difficulty with the interpolation of the advected
quantities is that advecting the volumetric densities tends to
generate unphysically high velocities in low cells with low mass density.  We
circumvent this problem by using consistent transport
\citep{Ston-Norm:92}, in which it is the specific
quantities that are interpolated, \ie
\begin{equation}
\lambda^*_{\pm i} = \overline{\rho}_{\pm i}
(\overline{\lambda/\rho})_{\pm i} \,,
\end{equation}
where $\overline{\rho}_{\pm i}$ and $(\overline{\lambda/\rho})_{\pm i}$
are the interpolated values of $\rho$ and the specific quantity $\lambda/\rho$
at the $\pm i$th boundary of the cell.

The choice of the method used for interpolating the advected quantities has
to be made carefully, so as to avoid introducing instabilities in the
finite-difference scheme.  Several such methods exist, of which we have
chosen to use upwinding methods. These methods provide stability by clipping
new local extrema, and limit diffusivity by interpolating quantities to
the boundary in a way that accounts for the difference between the
velocities associated with the upwind and downwind characteristics. Upwinding
methods of various orders exist, with the the higher-order methods being necessarily
more computationally expensive. The three methods we have implemented are the
donor cell (zeroth-order), van Leer (first-order), and piecewise parabolic
advection (PPA; second-order) methods.

\subsubsection{Donor Cell Upwinding} \label{DS:A:DCU}
The donor cell method is a zeroth-order upwinding scheme, approximating the
spatial distribution of a given quantity, $q$, as a step function.  In this
method, all information from the downwind cell is ignored, \ie~at the
$-i$th cell boundary
\begin{equation}
\overline{q}_{-i}
=
\left\{
\begin{aligned}
&q_{-i} & {\rm if}\quad \overline{v}_{-i} \ge 0 \\
&q & {\rm if}\quad \overline{v}_{-i} < 0
\end{aligned}
\right. \,.
\end{equation}
For a given cell, this only requires information from the nearest
neighbours.  In practise, donor cell upwinding is highly diffusive
(see, \eg,~\S\ref{TP:A}), and hence was not used beyond the testing
stage.

\subsubsection{van Leer Upwinding} \label{DS:A:vLU}
The van Leer upwinding method is a first-order method first described
by its namesake \citep{vanL:77a,vanL:77b,vanL:79}.  In contrast to the
donor cell method, the distribution of $q$ is approximated by a piecewise
linear function.  The slopes of these linear functions are given by
the so-called van Leer slopes, defined below for a given cell
along the $i$th direction,
\begin{equation}
dq^i = \left\{
\begin{aligned}
&\frac{ 2 (q_{+i}-q)(q-q_{-i}) }{\Delta x^i (q_{+i}-q_{-i})}
& {\rm if} \quad (q_{+i}-q)(q-q_{-i}) > 0 \\
&0 & {\rm otherwise}
\end{aligned}
\right. \,.
\end{equation}
In terms of the van Leer slopes, the upwinded value of the quantity
$q$ at the $-i$th cell boundary is given by
\begin{equation}
\overline{q}_{-i}
=
\left\{
\begin{aligned}
&q_{-i} + \frac12 \left( \Delta x^i - \overline{v}_{-i} \Delta t \right) dq^i_{-i}
& {\rm if} \quad \overline{v}_{-i} \ge 0 \\
&q - \frac12 \left( \Delta x^i + \overline{v}_{-i} \Delta t
\right) dq^i
& {\rm if} \quad \overline{v}_{-i} < 0
\end{aligned}
\right. \, ,
\end{equation}
where the notation $dq^i_{\pm j}$ denotes the van Leer slope in the
$i$th direction for the neighbouring cell in the $\pm j$th direction.
The van Leer method prevents the introduction of new local
extrema, and hence ensures stability in the advection scheme.
When the van Leer slopes vanish, the scheme reduces to the donor
cell method. Note that, because van Leer upwinding uses the van
Leer slopes of neighbouring cells, it requires information from both
the nearest and next-nearest neighbours.

\subsubsection{PPA Upwinding} \label{DS:A:PPAU}
The PPA method is a second-order upwinding method originally developed by
\citet{Cole-Wood:84}. It approximates the distribution of $q$
by a piecewise parabolic function. The essence of the method
is the determination of the monotonized left and right interface 
values, $q_L$ and $q_R$, which are computed via equations
(1.6)--(1.10) in \citet{Cole-Wood:84}. In terms of $q_L$ and $q_R$,
the upwinded value of $q$ at the $-i$th cell boundary is given by
\begin{equation}
\overline{q}_{-i}
=
\left\{
\begin{aligned}
&
\begin{aligned}
&q_{R,-i} + \xi(q_{-i} - q_{R,-i}) \\
& \ + \xi(1-\xi)(2q_{-i}-q_{R,-i}-q_{L,-i})
\end{aligned}
& {\rm if} \quad \overline{v}_{-i} \ge 0 \\
&
\begin{aligned}
&q_L + \xi(q - q_L) \\ & \ + \xi(1-\xi)(2q-q_R-q_L)
\end{aligned}
& {\rm if} \quad \overline{v}_{-i} < 0
\end{aligned}
\right. \,,
\end{equation}
where $\xi=\overline{v}_{-i}\Delta t/\Delta x^i$. This requires
information from the nearest three neighbours.

The PPA method is substantially less diffusive than the van Leer
method. This is especially notable at discontinuities, where the
profiles generated by PPA are significantly steeper than those
generated by the van Leer scheme.  However, the improvement comes with
a relatively high computational cost.  It has been found by
\citet{Ston-Norm:92} that, typically, increasing the grid resolution
is a computationally more efficient way to obtain greater
accuracy. For this reason, unless explicitly stated otherwise, we use
the van Leer upwinding method.

\subsection{Artificial Viscosity} \label{DS:AV}
In Eulerian upwinding schemes, shocks can lead to numerical instabilities.
If resolving shocks is critical, the instabilities may be cured via the
introduction of Riemann solvers (capable of localising a shock to a single
cell boundary).  However, if resolving shocks is unnecessary, it is significantly
easier to introduce an artificial numerical viscosity to smooth them out.
Several prescriptions for implementing numerical viscosity can be found
in the literature; we chose to implement the von Neumann-Richtmyer scheme
because of its ability to produce the correct shock propagation velocity and its
low dissipation far from shocks \citep[a direct result of the fact that
it acts only in regions of compression;][]{Ston-Norm:92}.
This scheme takes the form of defining a viscous pseudo-pressure for
each direction which is non-vanishing in regions of compression only:
\begin{equation}
Q^i = \left\{\begin{aligned}
&l^2 \rho \left(\frac{\partial v^i}{\partial x^i} \right)^2 
& {\rm if} \quad \frac{\partial v^i}{\partial x^i} < 0 \\
&0 & {\rm otherwise}
\end{aligned}
\right. \,,
\end{equation}
for $i=x,\,y,\,z$, where $l$ is the length scale over which shocks
are to be smoothed. The associated source term for equation
(\ref{momentum_cons}) is given by
\begin{equation}
F^i_{\rm visc} = -\frac{\partial Q^i}{\partial x^i} \,.
\end{equation}
Typically this will smooth a shock front over
a number of cells--a distance that is usually much larger than the
natural shock depth.
It should also be noted that a strictly correct treatment of shocks is
precluded by the adiabatic condition, equation (\ref{entropy_cons}).
This can be remedied by the inclusion of a viscous source term in the
entropy equation.  However, since for the applications we envision
shocks will result in the rapid thermalisation of the kinetic
energy of the stellar oscillations, their mere production may make a
purely hydrodynamic description inapplicable.  In particular,
thermonuclear processes could dominate at such a point, and thus
neither the added complexity and computational overhead of the Reimann
solver methods nor the complication of an entropy source term are required.

\subsection{Momentum Source Terms} \label{DS:MST}
In addition to advection, the momentum density evolves due to pressure
gradients, self-gravity, and external forces (if any).
We have found that simply finite-differencing $\bmath{\nabla} P$ leads to
a less stable system than calculating the gradient via partial derivatives
of the equation of state, and finite-differencing in $\rho$ and $s$.
In contrast, the gravitational acceleration is obtained directly in terms of a
second-order, finite-difference of the gravitational potential
(the details of solving for which are presented in \S\ref{StPE}).
The finite differencing of the viscous force is
performed in two steps: ({\em i}) determining the viscous
pseudo-pressure, and ({\em ii}) finite differencing the viscous
pseudo-pressure to obtain the viscous force directly.  In finite
difference form, the viscous pseudo-pressure is defined by
\begin{equation}
Q^i_{\pm i} = \left\{
\begin{aligned}
&\eta \frac{\rho_{\pm i} + \rho}{2}
\left(\frac{v^i_{\pm i} - v^i}{\Delta x^i} \right)^2 
& {\rm if} \quad \pm \left( v^i_{\pm i}-v^i \right) < 0 \\
&0 & {\rm otherwise}
\end{aligned}
\right. \,,
\end{equation}
for $i=x,\,y,\,z$.  The dimensionless coefficient $\eta$ is
approximately the number of cells over which discontinuities are
to be smoothed.  Typically, we find $\eta=2$ to be adequate.
The viscous force is then determined by
\begin{equation}
F^i_{\rm visc} = - \frac{Q^i_{+i}-Q^i_{-i}}{\Delta x^i} \,.
\end{equation}
Therefore, excluding external forces, the source terms in equation
(\ref{momentum_cons}) are given by
\begin{multline}
- \left( \frac{\partial P}{\partial \rho} \right)_s \frac{\rho_{+i} -
  \rho_{-i}}{2 \Delta x^i} -
\left( \frac{\partial P}{\partial s} \right)_\rho \frac{s_{+i} -
  s_{-i}}{2 \Delta x^i} \\
- \rho \frac{\Phi_{+i} - \Phi_{-i}}{2 \Delta x^i} 
+ F^i_{\rm visc}\,,
\label{fd_source_terms}
\end{multline}
for $i=x,\,y,\,z$.

When using a barotropic equation of state, $P(\rho)$, it can be
convenient to write the source terms in terms of the specific
enthalpy, $h$,
\begin{equation}
- \rho \left( 
\frac{h_{+i} - h_{-i}}{2 \Delta x^i}
+
\frac{\Phi_{+i} - \Phi_{-i}}{2 \Delta x^i}
\right)
+ F^i_{\rm visc}\,,
\label{enthalpy_fd_source_terms}
\end{equation}
for $i=x,\,y,\,z$.  An example of when this is useful will be
discussed in \S\ref{AtaPWD}.  Note that in this case, the entropy
equation is superfluous.

\subsection{Courant-Friedrichs-Lewy Time Step} \label{DS:CFLTS}
The stability of our explicit finite-difference scheme requires that the
time step should satisfy the Courant-Friedrichs-Lewy (CFL) criterion.
This corresponds to the physical consideration that, in a single time step,
information should only propagate into a given cell from the neighbouring cells
which are used to compute spatial derivatives at that point.
A time step that is too large would require information from more distant
cells, which is not available in the differencing scheme.
Therefore, for stability,
\begin{equation}
\Delta t \le t_{\rm CFL} \,,
\end{equation}
where the CFL time is defined by
\begin{equation}
t_{\rm CFL}
=
\min
\left(
\frac{\Delta x}{c_s + |v^x|},\,
\frac{\Delta y}{c_s + |v^y|},\,
\frac{\Delta z}{c_s + |v^z|}
\right) \,,
\label{CFL_time}
\end{equation}
where $c_s$ is the local adiabatic sound speed
\citep[see \eg][and references
therein]{Motl-Tohl-Fran:02,Ston-Norm:92}.
In addition, the inclusion of an artificial viscosity imposes the
additional requirement that the time step does not exceed the
timescale for diffusion across cell width length-scales:
\begin{equation}
t_{\rm visc} = \min \left(
\frac{\Delta x}{4 \eta |\Delta v^x|},
\frac{\Delta y}{4 \eta |\Delta v^y|},
\frac{\Delta z}{4 \eta |\Delta v^z|}
\right) \,,
\label{viscous_time}	
\end{equation}
\citep[see \eg][]{Ston-Norm:92}.
In practise, for many operator split methods, taking the time step to
be the CFL time does not ensure stability. Rather, it is necessary to
take $\Delta t$ to be some fraction of $t_{\rm CFL}$ or
$t_{\rm visc}$.  In practise, we find that a robust choice is
\begin{equation}
\Delta t \le \frac14 \min \left( t_{\rm CFL}, t_{\rm visc} \right) \,.
\end{equation}

From equation (\ref{CFL_time}) it is clear that the cells with the highest
velocities (both kinetic and sound) will provide the most stringent limits
on the time step.  An example is the case of cells constituting the
vacuum surrounding a star.  In practise, for numerical reasons,
no portion of the grid can have vanishing mass density.
Therefore, we take `zero' density to be some small fraction (typically, $10^{-8}$)
of the initial maximum density.  As a result, the vacuum is physically insignificant.
Nonetheless, because of their large accretion velocities (though negligible momentum
densities), the vacuum cells can be the limiting factor in determining the time
step.  To avoid this problem, we impose a velocity cap, so that the CFL time is set
by only considering cells with densities larger than, say, $10^{-6}$ of the maximum
density.\footnote{What is important is that the density cut-off used for the CFL
time is large enough to exclude the vacuum cells.} The remaining cells have their
velocities capped at
\begin{equation}
v_{\rm cap}
=
\min
\left(
\frac{\Delta x}{\Delta t},\,
\frac{\Delta y}{\Delta t},\,
\frac{\Delta z}{\Delta t}
\right)
\,,
\end{equation}
so as to not drive the time step down.  While this explicitly violates
the hydrodynamic equations presented in \S\ref{GHE}, it does so in a
physically negligible manner.

We use operator splitting to separate the source and advection
contributions to the evolution of the fluid quantities at each time
step.  However, we do not use directional splitting,
making our scheme a variation of the unsplit method of van Leer.
Thus, a single time step is taken in three stages: (1) taking half of
the source step, (2) performing the updates due to advection, and (3)
taking the second half of the source step.  The gravitational
potential is calculated at each source sub-step.

\subsection{Boundary Conditions} \label{DS:BC}
Because the upwinding methods require information about neighbouring
cells, it is necessary to provide a boundary of ghost cells along the
outer edges of the grid.  As these ghost cells are not evolved
themselves, they require some prescription for assigning the evolved
quantities to them.  We have implemented three types of boundary conditions:
fixed, replicated, and outflow.

The first, and simplest, is the fixed boundary condition.  In this
prescription, the boundary cells are fixed to have `zero' density,
entropy density, and momentum flux.  This tends to limit the velocity
of the `zero' density vacuum by not providing a boundary momentum
flux.

The second set of boundary conditions consists of replicating
the last set of cells in the grid.  This provides a slightly more
realistic set of boundary conditions, allowing the accretion of the
`zero' density vacuum to stabilise through hydrodynamic balance.
However, if a physically significant portion of the flow is crossing
the boundary, then this is significantly superior to the first scheme.

The third set of boundary conditions implemented are the so-called
outflow boundary conditions.  In this prescription, fluid is allowed
to flow off the grid but not into it.  In order to prevent the
boundaries from physically affecting the fluid on the grid, the
boundary values for density and entropy are chosen to preserve
hydrostatic equilibrium in the last grid zone.  Note that this does
not stop the fluid from advecting off the grid through this zone.
As a result, this will minimise the creation of spurious
reflections at the boundaries.  For a self-gravitating fluid configuration
that is initially contained entirely within the grid, this provides the most
realistic set of conditions.

\subsection{Parallelisation} \label{DS:P}
The primary purpose for the development of our code is to perform high
resolution studies of the non-linear evolution of normal modes in
white dwarfs.  The resulting computational requirements necessitate
high-performance computing.  Because the sourced advection step for a
given cell depends only upon cells in its immediate neighbourhood, it
naturally lends itself to a straightforward parallelisation scheme.
This takes the form of dividing the entire grid into a number of
sub-domains, each of which are handled by a separate process.
Because interprocess communication incurs substantial performance
penalties, we need to choose a domain decomposition that minimises the
communication required.  The source of interprocess communication in
each sourced advection step is the need for neighbour data around the
edges of each sub-domain.  Therefore, the time penalties due to
interprocess communication are dictated by the surface area of each
sub-domain, as well as the depth of neighbours that is necessary (one
for donor cell upwinding, two for van Leer upwinding, and three for PPA
upwinding).  Hence, minimising the surface area of each sub-domain
minimises the interprocess communication.

We have chosen to implement our code in the C++ programming language.
This choice is motivated by considerations such as modularity of
design, flexibility, efficiency, ease of code reuse, and
extensibility.  For example, using the object-oriented paradigm in the C++
language has allowed us to maintain a clean separation between
interfaces and implementations (\eg,~for the equation of state,
Poisson equation solver, and initial conditions \etc), and features
such as templates have allowed us to write generic code without
sacrificing runtime performance.

Since standard C++ does not provide facilities for parallel computing,
it is necessary to use additional libraries to handle the
parallelisation.  We have chosen to implement parallelisation via the
Message Passing Interface (MPI).  Since both optimising, ISO-compliant
C++ compilers and high quality MPI implementations are available for
virtually every major computing platform, our code is highly portable.

\section{Solving the Poisson Equation} \label{StPE}
Equation (\ref{poisson_eq}) is distinct from equations
(\ref{mass_cons}-\ref{momentum_cons}) in that it
requires global, rather than local, information.
There are a number of methods that can be used to solve the Poisson equation.
These include general elliptic equation set solvers, multigrid methods, multipole
methods, and spectral methods
\citep[see \eg][]{Motl-Tohl-Fran:02,Flash:00,Mull-Stei:95,Ston-Norm:92}.
Spectral methods tend to be the most efficient, and implementing them
on a regular Cartesian grid is straightforward.

The solution of the Poisson equation requires the specification of a
boundary condition on some closed surface. In most physical problems, this
surface is chosen to lie at infinity, upon which the potential is chosen
to vanish. However, since our computational domain is finite,
it is not possible to impose a boundary condition at infinity in a
straightforward manner. Instead, we define the value of the potential
on the surface of our domain, which we compute via a multipole expansion:
\begin{equation}
\Phi^B({\bf x}) = -\sum_{l=0}^{\infty}\sum_{m=-l}^{l}
                  \frac{4\pi G}{2l+1} r^{-l-1} Q_{lm} Y_{lm}({\bf \hat{x}}) \,,
\label{multipole_eq}
\end{equation}
where
\begin{equation}
Q_{lm} = \int d{\bf x}' r'^l Y_{lm}^*({\bf \hat{x}}') \rho({\bf x}') \,.
\end{equation}
In practise, it is only necessary to include the first few multipoles
(for our purposes $l_{\rm max} = 5$) to obtain accurate boundary values.
Note that the boundary condition at infinity is built into the multipole expansion.

Given the Dirichlet boundary condition, it is possible to solve
Poisson equation via a discrete sine transform (DST)
\citep[see \eg][]{Pres-Teuk-Vett-Flan:92}.  Written in its
finite-difference form, (\ref{poisson_eq}) becomes
\begin{equation}
\sum_{i=x,y,z} \frac{\Phi_{+i} - 2 \Phi + \Phi_{-i}}{(\Delta x^i)^2}
= 4 \pi G \rho \,.
\label{fd_poisson_eq}
\end{equation}

In terms of their discrete sine transforms $\widehat{\Phi}$ and
$\widehat{\rho}$, $\Phi$ and $\rho$ are given by
\begin{align}
\label{phi_dst_eq}
\Phi_{i,j,k} &= \frac{2}{I J K}
                \sum_{m=1}^{I-1} \sum_{n=1}^{J-1} \sum_{p=1}^{K-1}
                \widehat{\Phi}_{m,n,p}
                \sin\frac{\pi i m}{I} \sin\frac{\pi j n}{J} \sin\frac{\pi k p}{K} \\
\rho_{i,j,k} &= \frac{2}{I J K}
                \sum_{m=1}^{I-1} \sum_{n=1}^{J-1} \sum_{p=1}^{K-1}
                \widehat{\rho}_{m,n,p}
                \sin\frac{\pi i m}{I} \sin\frac{\pi j n}{J} \sin\frac{\pi k p}{K} \,,
\end{align}
where $i$, $j$, $k$, and $I$, $J$, $K$ define the location in, and the
dimensions of, the computational domain, respectively.
Substituting these expansions into (\ref{fd_poisson_eq}) gives
\begin{equation}
\label{phihat_from_rhohat_eq}
\widehat{\Phi}_{m,n,p} = -4\pi G\frac{\widehat{\rho}_{m,n,p}}{\kappa^2_{m,n,p}} \,,
\end{equation}
where
\begin{equation*}
\begin{aligned}
\kappa^2_{m,n,p} &= \frac{2}{(\Delta x)^2}\left( 1 - \cos\frac{\pi m}{I} \right) \\
                 &+ \frac{2}{(\Delta y)^2}\left( 1 - \cos\frac{\pi n}{J} \right) \\
                 &+ \frac{2}{(\Delta z)^2}\left( 1 - \cos\frac{\pi p}{K} \right) \,.
\end{aligned}
\end{equation*}
The potential $\Phi_{i,j,k}$ is then computed from (\ref{phi_dst_eq}).

Expanding $\Phi$ in terms of the sine basis functions of the Fourier
series ensures that it vanishes at the boundaries of the domain.
Non-zero boundary conditions can be incorporated by adding an
appropriate source term to the right side of equation
(\ref{fd_poisson_eq}).  We may define $\Phi' = \Phi - \Phi^B$
where now $\Phi^B$ is determined by equation (\ref{multipole_eq}) at
one zone beyond the boundary and vanishes everywhere else.  The
resulting equation for $\Phi'$ is the same as equation
(\ref{fd_poisson_eq}) in the interior and is given by
\begin{equation}
\sum_{i=x,y,z}
\frac{\Phi'_{+i} - 2 \Phi' + \Phi'_{-i}}{(\Delta x^i)^2}
= 4 \pi G \rho - \frac{\Phi^B_{\pm j}}{(\Delta x^j)^2}
= 4 \pi G \rho' \,,
\label{fd_poisson_eq2}
\end{equation}
on the $\pm j$th boundary.  As a result, the effective source terms
are given by
\begin{equation}
\label{eff_source_eq}
\begin{aligned}
4\pi G\rho'_{i,j,k} &= 4\pi G\rho_{i,j,k} \\
&- \frac{1}{(\Delta x)^2}\left(\delta_{i,1}\Phi^B_{0,j,k} + \delta_{i,I-1}\Phi^B_{I,j,k}\right) \\
&- \frac{1}{(\Delta y)^2}\left(\delta_{j,1}\Phi^B_{i,0,k} + \delta_{j,J-1}\Phi^B_{i,J,k}\right) \\
&- \frac{1}{(\Delta z)^2}\left(\delta_{k,1}\Phi^B_{i,j,0} + \delta_{k,K-1}\Phi^B_{i,j,K}\right) \,.
\end{aligned}
\end{equation}

To summarise, our procedure for solving the Poisson equation is:
\begin{enumerate}
\item Calculate $\Phi^B$ via the multipole expansion (\ref{multipole_eq}).
\item Calculate the effective source terms for $\Phi'$ from (\ref{eff_source_eq}).
\item Perform a DST on the effective source terms.
\item Calculate $\widehat{\Phi}'$ from (\ref{phihat_from_rhohat_eq}).
\item Perform a DST on $\widehat{\Phi}'$ to determine $\Phi'$.
\end{enumerate}
We do not actually need to add $\Phi^B$ to our final answer since it only
affects the ghost points outside our grid.  Note that, because we use a
second-order finite-difference to determine the gravitational acceleration in
equation (\ref{fd_source_terms}), it is necessary to define $\Phi$ on
an extra surface of ghost cells on each edge of the domain.

The DST is most efficiently parallelised in terms of a slab decomposition
of the grid, as opposed to the ideal decomposition for the sourced
advection step (which is cubical). As a result, a significant amount of
interprocess communication is required to prepare for the solution of the
Poisson equation at each source sub-step. However, we have found that the time
saved by using the DST more than outweighs the penalty incurred by the
communication overhead compared to alternative methods.

\section{Test Problems} \label{TP}
\subsection{Advection} \label{TP:A}
\begin{figure}
\begin{center}
\includegraphics[width=\columnwidth]{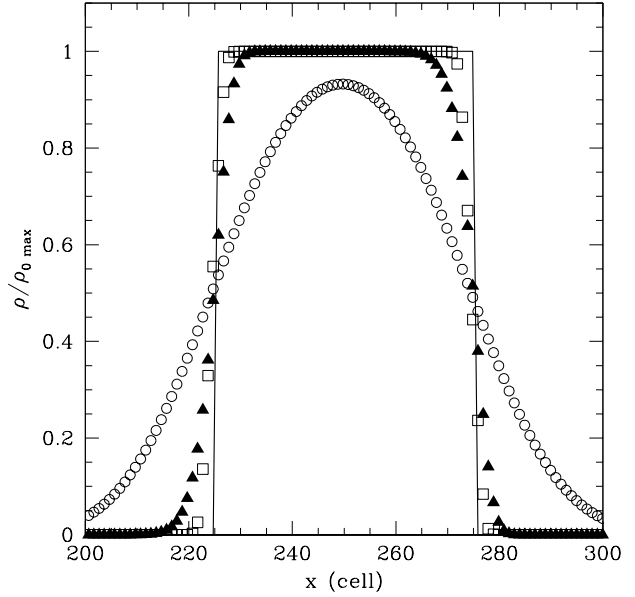}
\end{center}
\caption{
A square pulse that has been advected five times its initial width (50 cells)
using the donor cell (open circles), van Leer (filled triangles), and PPA
(open squares) upwinding schemes.  For reference, the original pulse profile
is also shown.
}
\label{advect}
\end{figure}
In order to test the advection scheme, we considered the advection of a
square pulse (without source terms).  In Figure \ref{advect}, the
pulse is shown after being advected five times its initial width (50
cells) using both the donor cell and van Leer upwinding methods.  It
is clear that both methods are diffusive, with the donor cell method
substantially more so.

\begin{figure}
\begin{center}
\includegraphics[width=\columnwidth]{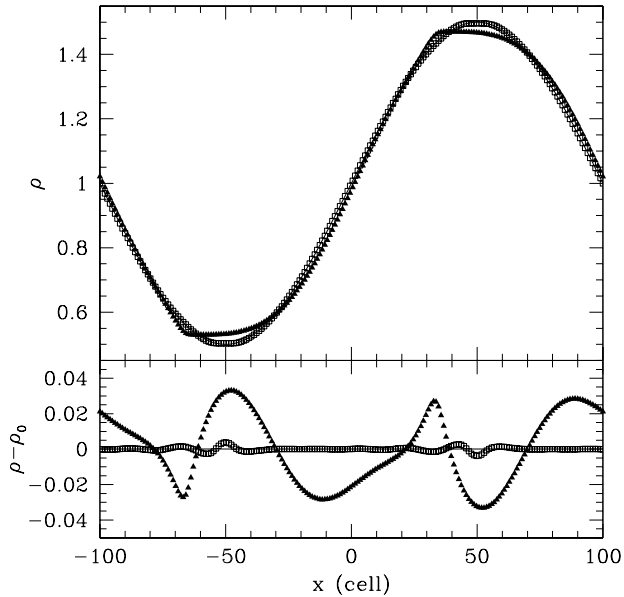}
\end{center}
\caption{A sine wave is advected with periodic boundary conditions for
  100 times its wavelength (200 cells) using the van Leer (filled
  triangles) and PPA (open squares) upwinding schemes.  In the top
  panel the density profile is shown explicitly, while in the bottom the
  residuals are plotted.  For reference the analytical result is also shown.}
\label{sine_wave}
\end{figure}
In general, diffusion will lead to errors in both the amplitude and the
phase of an advected pulse.  In order to quantify these errors for
diffusion resulting from the upwinding scheme, a sine wave was advected
with periodic boundary conditions for 100 times its wavelength.  By
this time, the donor cell upwinding scheme has diffused the sine wave
completely, hence only the van Leer and PPA methods are shown in
Figure \ref{sine_wave}.  The errors are at the 4\% and 0.4\%
levels, respectively, with deviations becoming most significant at
extrema.  In both the square pulse and the sine wave, a noticeable
asymmetry (which is determined by the direction of propagation)
develops as a result of higher-order effects in the upwinding schemes.

\subsection{Sod Shock Tube} \label{TP:SST}
The pressure source term in equation (\ref{momentum_cons}) was tested
by the Sod shock tube problem.  The Sod shock
tube consists of an initial density and pressure discontinuity, and its
subsequent evolution for an ideal gas ($\Gamma = 1.4$) and a specific
set of initial conditions.  For $x>0$, $\rho=0.125$
and $P=0.1$, while for $x \le 0$, $\rho=1$ and $P=1$.  Because it is the
entropy density and not the pressure that is evolved, it is necessary to
find $s$ as a function of $\rho$ and $P$ for an ideal gas:
\begin{equation}
s = \ln \left( \frac{P^n}{\rho^{n+1}} \right)
\quad
{\rm where}
\quad
n = \frac{1}{\Gamma-1} \,.
\end{equation}
The Sod shock tube is useful as a test because the resulting $\rho$
and $P$ profiles for any given time can be calculated analytically
\citep[see, \eg,][]{Sod:78,Hawl-Wils-Smar:84}.
\begin{figure}
\begin{center}
\includegraphics[width=\columnwidth]{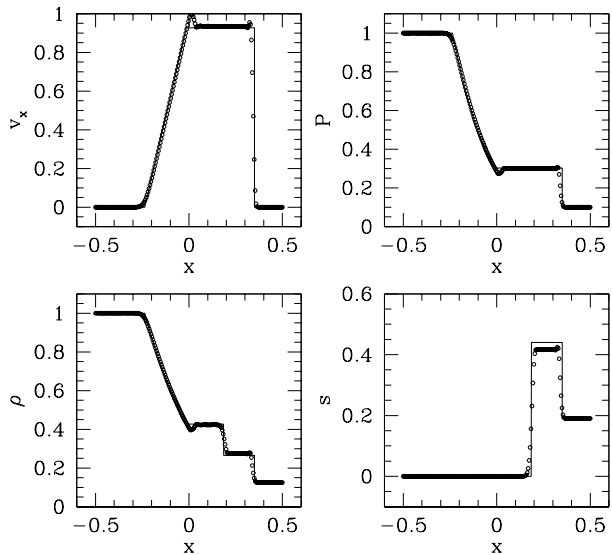}
\end{center}
\caption{
The density, pressure, velocity, and entropy are shown for the Sod shock
tube at $t=0.2$ (the units of which depend upon the units chosen for the
pressure and density).  200 cells were used with van Leer upwinding.
The head and tail of the rarefaction wave are located at $x=-0.235$ and
$x=-0.014$, respectively.  The contact and shock discontinuities are at
$x=0.184$ and $x=0.35$, respectively.
}
\label{sod}
\end{figure}

In Figure \ref{sod}, the numerical results from our code are compared
to the analytical solutions.  Overall, they are in good agreement,
with the exception of two minor discrepancies.
The most notable discrepancy is the entropy deficit in
the post-shock fluid ($0.184<x<0.35$).  This is a result of using the
adiabatic condition, and thus ignoring entropy production at shocks.
Hence, the higher analytical value is easy to understand.  Because we
intend to apply the code to scenarios in which the adiabatic condition
holds to a very good approximation, we expect the entropy deficit to be
physically insignificant.  The second discrepancy is the overshoots at
points where the slopes of quantities change discontinuously.  As
discussed in \citet{Ston-Norm:92}, this is a real result, originating
from the numerical viscosity inherent in any finite-difference code.
The most important result, however, is the fact that the artificial
viscosity causes the shock fronts to be well behaved in our code.

\subsection{Pressure-Free Collapse} \label{TP:PFC}
\begin{figure}
\begin{center}
\includegraphics[width=\columnwidth]{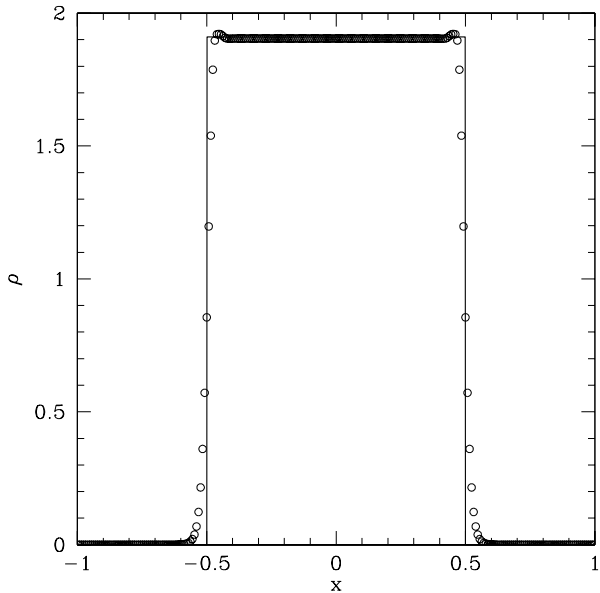}
\end{center}
\caption{
The numerical (open circles) and analytical (solid line) solutions for
the density as a function of distance along a radial section for 
the pressure free collapse of a uniform density sphere are shown.
The initial radius and total mass of the sphere was unity.
A $256\times256\times256$ cell grid was used.  With
Newton's constant given by $G=1$, this occurs at $t=0.909$. 
}
\label{freefall}
\end{figure}

The gravitational source term in equation (\ref{momentum_cons}) was
tested via the pressure-free collapse of a uniform density sphere.
Once again, there is an analytical solution:
\begin{align}
r &= r_0 \cos^{2} \beta \nonumber \\
\rho &= \rho_0 \cos^{-6} \beta \\
t &= \left( \beta + \frac12 \sin 2 \beta \right)
\left( \frac{8\pi}{3} G \rho_0 \right)^{-1/2}
\nonumber \,,
\end{align}
\citep[see, \eg,][]{Ston-Norm:92}.
Figure \ref{freefall} depicts the result after allowing the radius to
halve (at $t=0.909$ for $G=1$), for a $256\times256\times256$ cell
grid.  There is a small excess on the edges resulting from our
implementations of viscosity and consistent transport (which
necessarily treats the advection of velocity into the edges
differently due to the density gradients).  Overall, it does show good
agreement with the analytical prediction.

\section{Application to a Pulsating White Dwarf} \label{AtaPWD}
\subsection{Hydrostatic Equilibrium} \label{AtaPWD:HE}
The problem of choosing an equilibrium fluid configuration is made
non-trivial by the finite differencing of the the dynamical
equations.  Consequently, a method to produce an equilibrium
solution for the finite difference equations is required.
For a barotropic equation of state, we have chosen to make use of the
self-consistent field (SCF) method
\citep[see, \eg,][]{Motl-Tohl-Fran:02,Hach:86,Ostr-Mark:68}.
Because it is well described elsewhere, we will only summarise the
procedure here.
\begin{enumerate}
\item An initial guess for the density (taken from the continuous
  solution) is used to generate the gravitational potential via the
  method described in \S\ref{StPE}.
\item The new gravitational potential and the initial density guess
  are then used to calculate the Bernoulli constant at the centre of
  the star.
\item the Bernoulli constant and the new gravitational potential are
  used to calculate the enthalpy at all points on the grid, which is
  then subsequently inverted to yield the new density guess.
\end{enumerate}
This procedure is iterated until the Bernoulli constant converges to
some specified tolerance--\ie~when the fractional change is less than
some small value (say, $10^{-12}$).  The resulting density
distribution is a solution to
\begin{equation}
\frac{h_{+i} - h_{-i}}{2 \Delta x^i}
+
\frac{\Phi_{+i} - \Phi_{-i}}{2 \Delta x^i} = 0 \,,
\end{equation}
and, hence, no net momentum flux is generated if the source terms are given
by equation (\ref{enthalpy_fd_source_terms}).  Note that, if the source
terms are given by equation (\ref{fd_source_terms}), this {\em may} still
produce a net momentum flux, and is not necessarily a good
approximation to equilibrium in that case.

\begin{table}
\begin{center}
\begin{tabular}{cccccc}
\hline
\hline
Model & $M\,(M_\odot)$ & $R\,(10^6{\rm m})$ & $\omega_*\,({\rm Hz})$
& $\omega_{f2}\,({\rm Hz})$ & $\omega_{p2}\,({\rm Hz})$\\
\hline
CWD & 0.632 & 8.56 & 0.365 & 0.562 & 1.15\\
HWD & 0.632 & 11.2 & 0.243 & 0.560 & 0.749\\
\hline
\end{tabular}
\end{center}
\caption{Stellar properties for a cold white dwarf with (CWD) and
  without (HWD) an isothermal envelope.  Specifically, the mass,
  radius, fiducial stellar frequency $\omega_*=\sqrt{GM/R^3}$,
  frequency of the adiabatic quadrupolar fundamental mode, and the
  frequency of the lowest order adiabatic quadrupolar $p$-mode.  Note
  that the inclusion of the isothermal envelope does not change the
  mass appreciably while significantly increasing the radius.}
\label{WD_props}
\end{table}
\begin{figure}
\begin{center}
\includegraphics[width=\columnwidth]{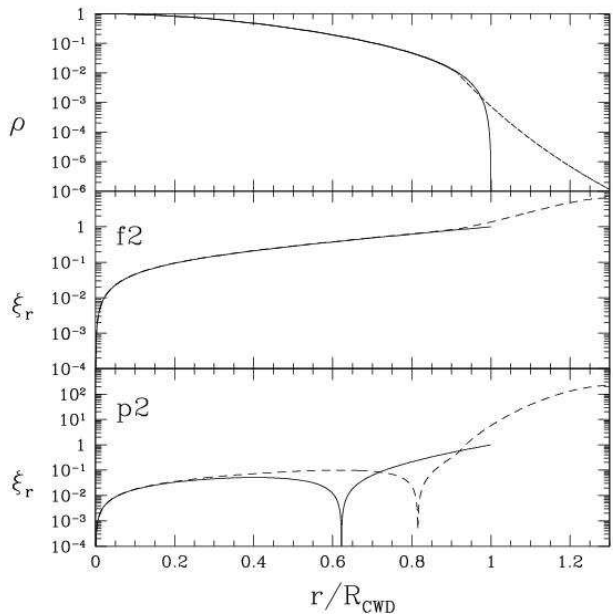}
\end{center}
\caption{Shown in the top panel are the density profiles for the cold
  white dwarf with (solid) and without (dashed) the isothermal
  envelope.  The two lower panels are the radial displacement profiles
  for the quadrupolar fundamental mode ($f2$) and the lowest order
  quadrupolar $p$-mode ($p2$) for the two models.  Note that the
  density and $f2$ mode profiles are very nearly the same for the two
  cases.  However, the mode profiles differ substantially for the $p2$
  mode.}
\label{models_compare}
\end{figure}
When
\begin{equation}
\left|\frac{\partial P}{\partial x^i}\right| > \frac{P}{\Delta x^i} \,,
\end{equation}
the pressure gradient required to preserve hydrostatic equilibrium
cannot be resolved on the grid.  For a star, this can result in
strong, inwardly directed forces at the surface, driving shocks into
the interior.  We have found that adding an isothermal envelope can
mitigate this problem by pushing the region where this inequality is
true off the grid, while adding an insignificant amount of mass to
the star itself.  This is done explicitly by setting a fiducial
density (which we chose to be $10^{-2}$ of the central density) at which
the equation of state changes from that of a cold white dwarf to a
$\Gamma=1$ polytrope.  The polytropic constant is chosen such that
$P(\rho)$ remains continuous across the transition.  Table
\ref{WD_props} compares the properties of the cold white dwarf with
(HWD) and without (CWD) the isothermal envelope.  Note that while the
isothermal envelope increases the radius significantly, it does not
change the mass or the frequency of the quadrupolar fundamental mode
($\omega_{f2}$).  The reason for this can be seen in Figure
\ref{models_compare}.  The $f2$ mode is more strongly weighted in the
core where the addition of the isothermal envelope makes no
difference.  In contrast, the lowest-order quadrupolar $p$-mode is
substantially affected by the presence of the envelope.  This probably
results from the fact that the radial wavelength of the $p2$ is
much closer to the height of the isothermal envelope.  Henceforth, all
evolutions were begun with the HWD model listed in Table
\ref{WD_props}.

\begin{figure}
\begin{center}
\includegraphics[width=0.8\columnwidth]{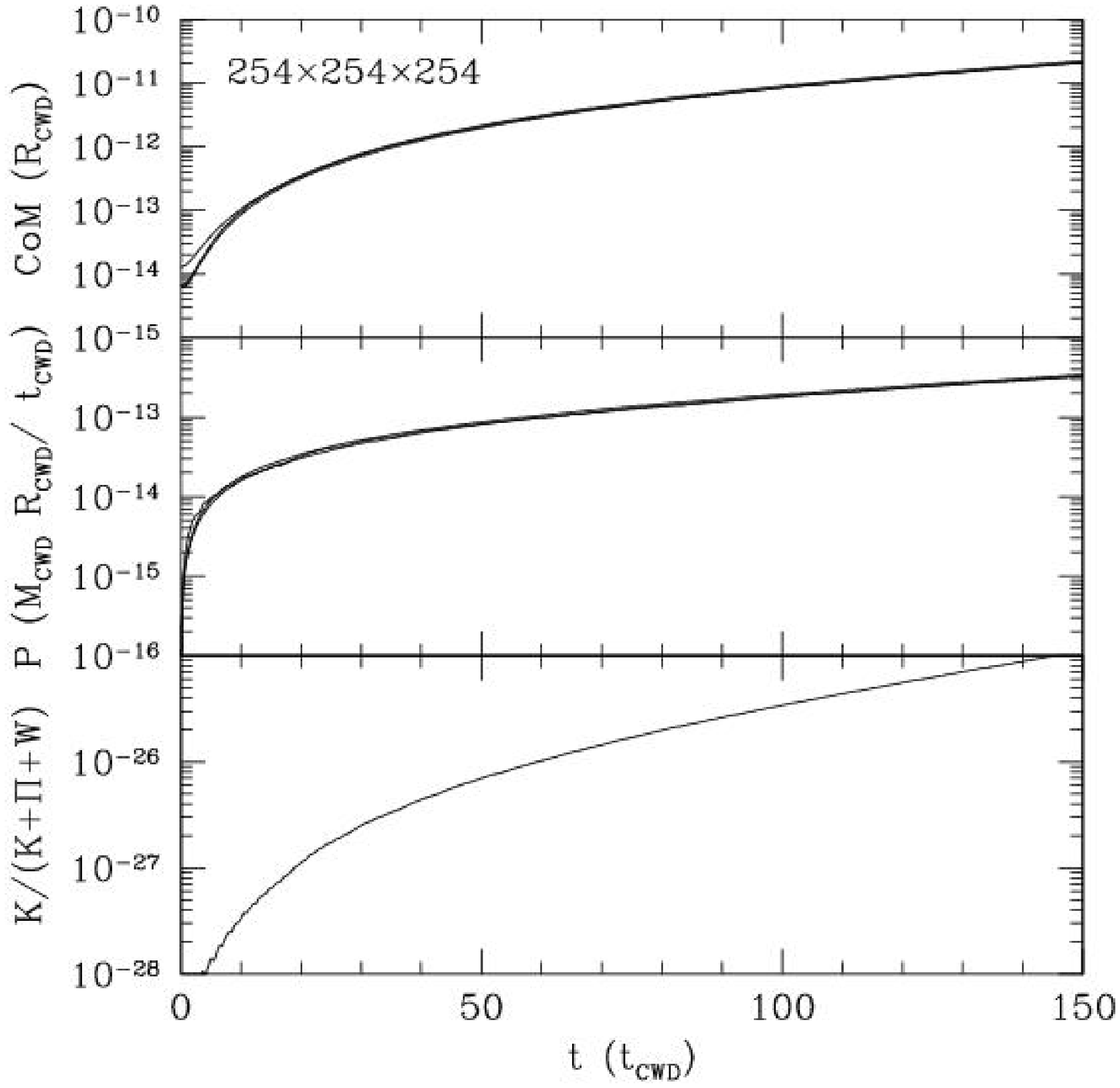}
\includegraphics[width=0.8\columnwidth]{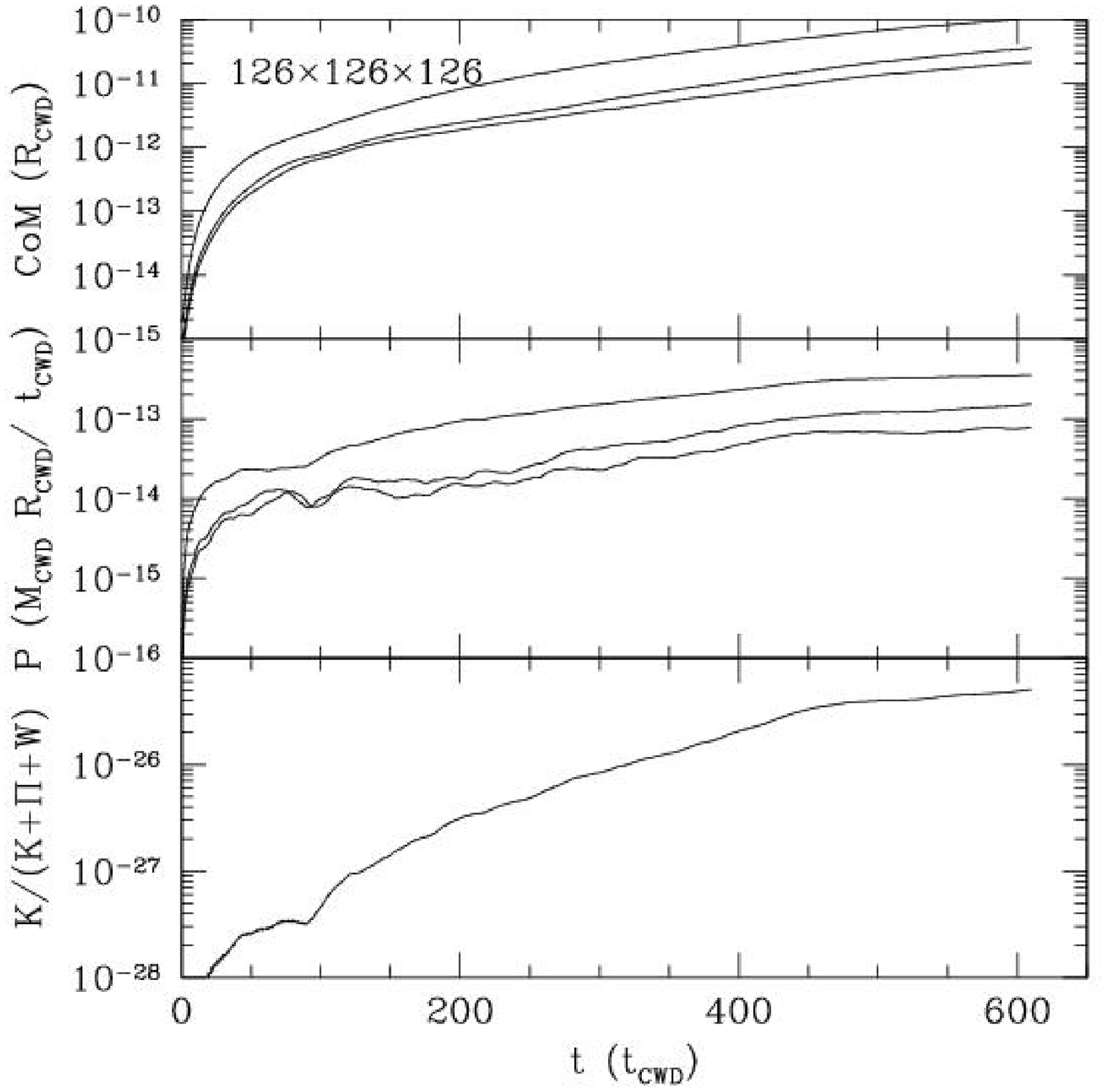}
\includegraphics[width=0.8\columnwidth]{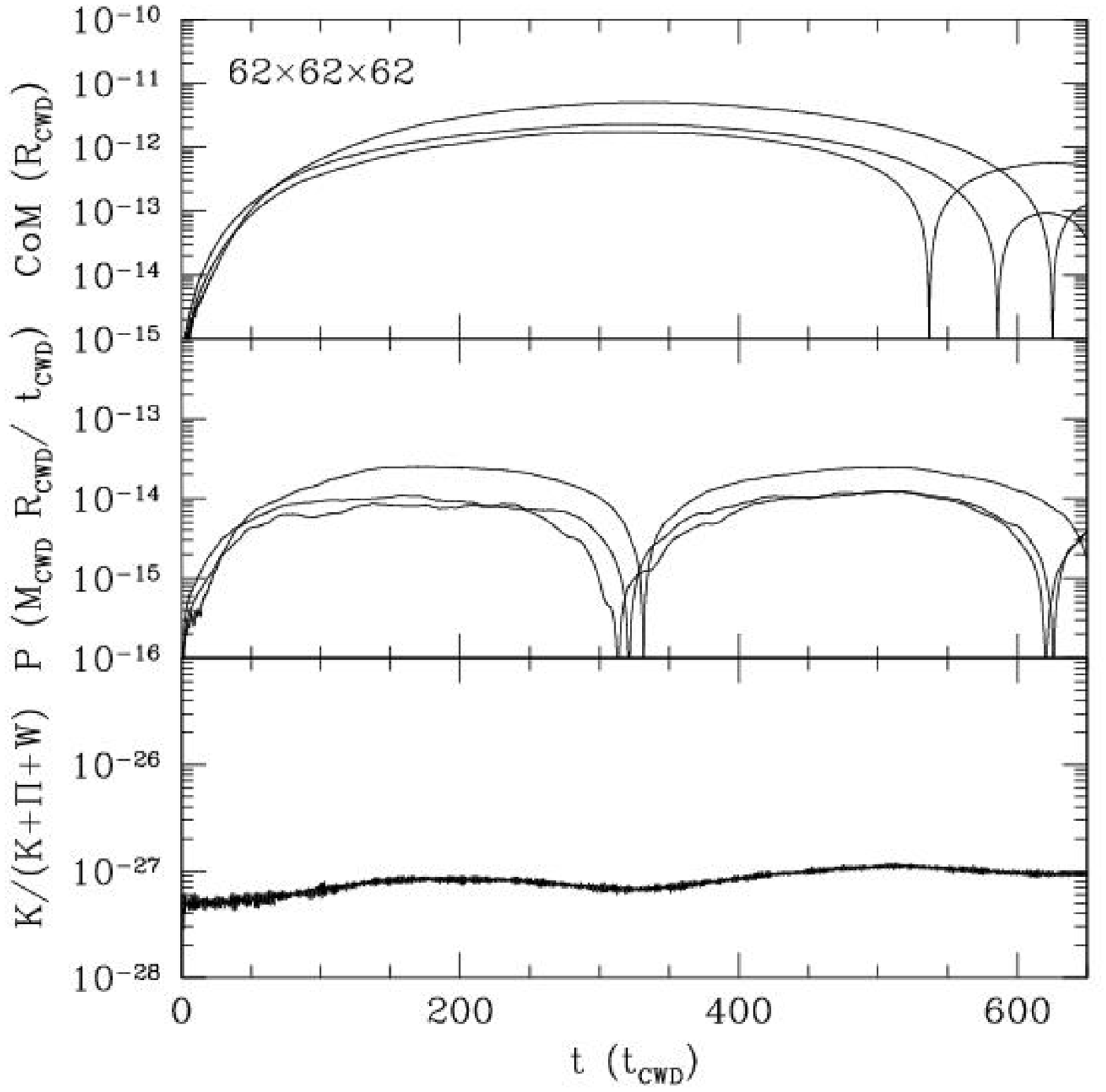}
\end{center}
\caption{Shown are the centre-of-mass (top panels), net momentum
  (middle panels), and fraction of the total energy converted into
  kinetic energy (bottom panels) for a number of grid resolutions
  (note the different time scales).  In all
  cases these quantities saturate well below significant levels (\eg,
  for the worst case, the centre-of-mass moves by less than $10^{-8}$
  cell widths in the 150 dynamical times shown, thus it would require
  roughly $10^{13}$ dynamical times before the centre-of-mass moves
  one stellar radius.  Typically, these appear to turn over, implying
  that they may never rise significantly above $10^{-7}$ cell widths.)
}
\label{sstars}
\end{figure}
The quality of the equilibrium generated by the SCF method may be
explicitly demonstrated.  Figure \ref{sstars} shows the evolution of the
centre-of-mass position, net momentum, and the fraction of the total
energy that is converted into kinetic energy for a star initially in
hydrostatic equilibrium.  The last quantity is given in terms of the
kinetic, internal, and gravitational components:
\begin{equation}
K=\int\frac{1}{2}\rho v^2\, d^3x\,,\quad
\Pi=\int p\, d^3x\,,\quad
W=\int\rho\Psi\, d^3x\,.
\end{equation}
Despite an initial exponential rise, these quantities saturate at
relatively low levels for all resolutions shown.  Note that all times
are measured in dynamical times of the cold white dwarf,
$t_{\rm CWD}\equiv 1/\omega_*$, which is approximately the time it
takes for a disturbance to cross the star.

\subsection{Oscillation Modes}
In general, the problem of interest is dynamical.  Specifically, we are
interested in the non-linear evolution of the oscillation modes of
a cold white dwarf which are being excited resonantly by tidal forces.
Towards this end, it is important to obtain a measure of the numerical
quality factor ($Q$; the $e$-folding time of the energy in the oscillation),
and the oscillation frequencies themselves.  That the latter may be
different from the frequencies in Table \ref{WD_props} is a result of
both the approximation of discrete cells {\em and} the fact that the
finite-difference equations are distinct from the continuous equations.
However, we expect the deviation to be small, and therefore a close
agreement between the predicted and observed frequencies serves as yet
another test for the correctness of our code.  Both the quality factor
and the oscillation eigenfrequencies can be obtained by deforming the star
in a particular way, and analysing the subsequent oscillations.

\begin{figure}
\begin{center}
\includegraphics[width=0.8\columnwidth]{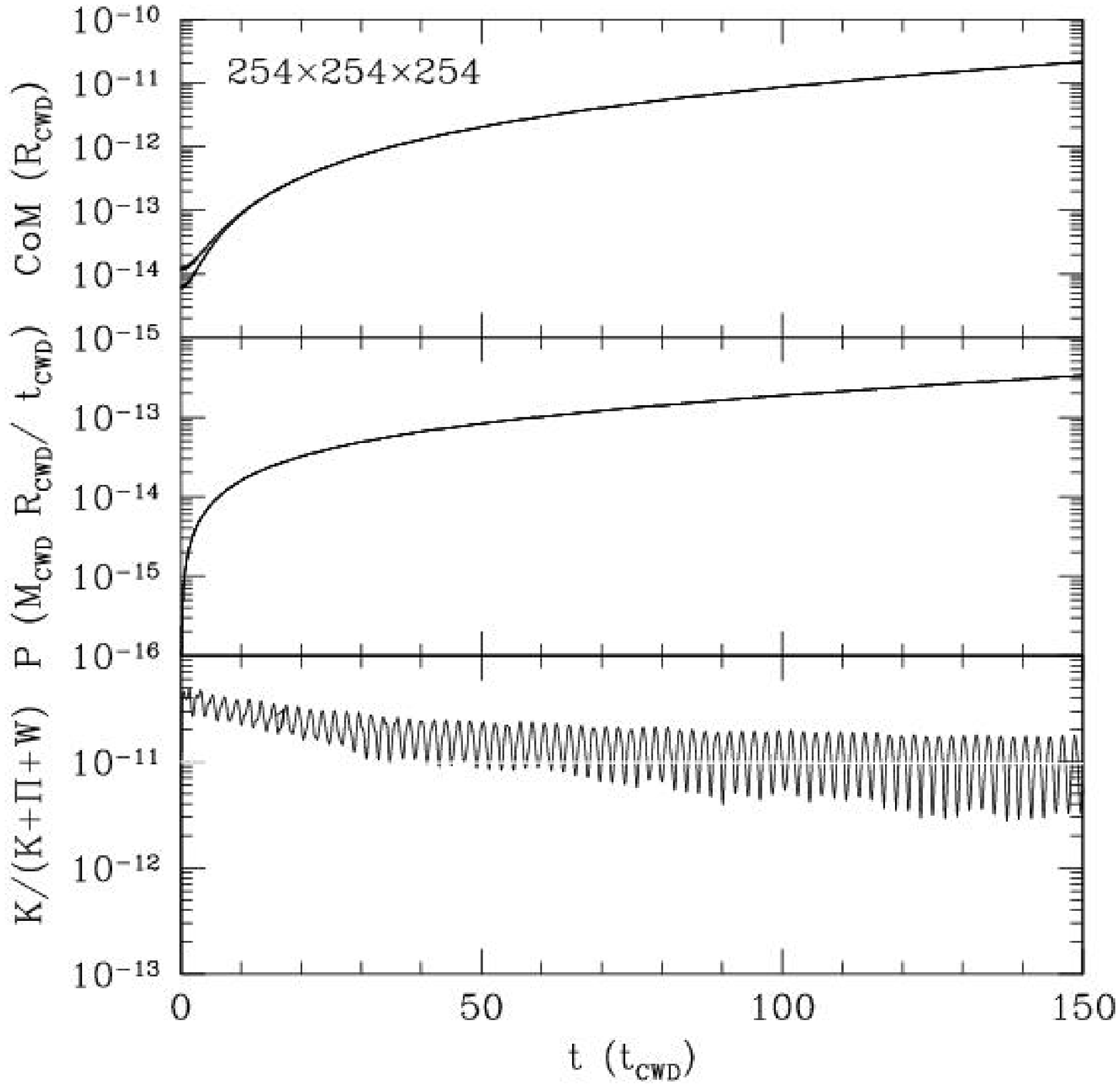}
\includegraphics[width=0.8\columnwidth]{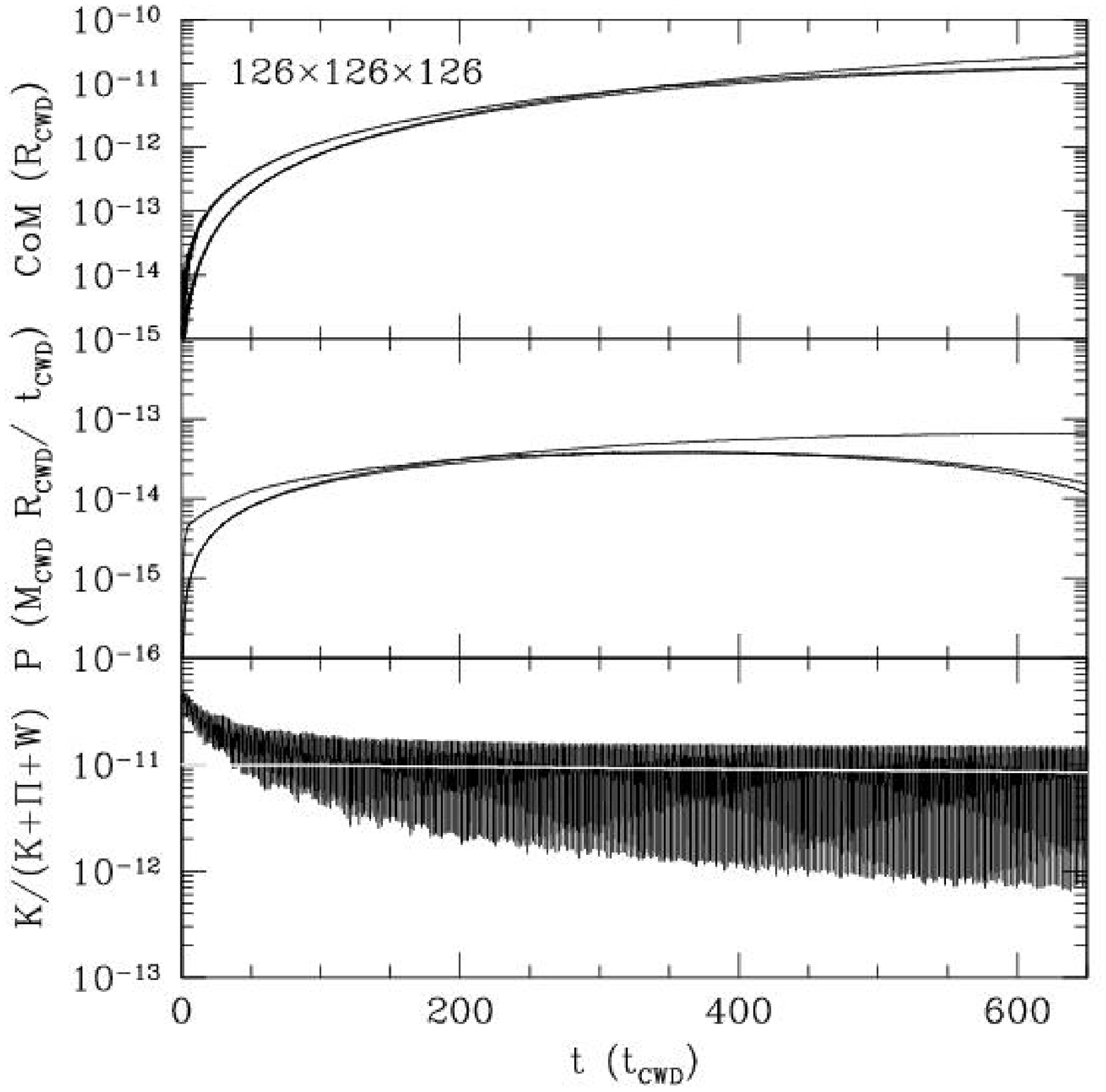}
\includegraphics[width=0.8\columnwidth]{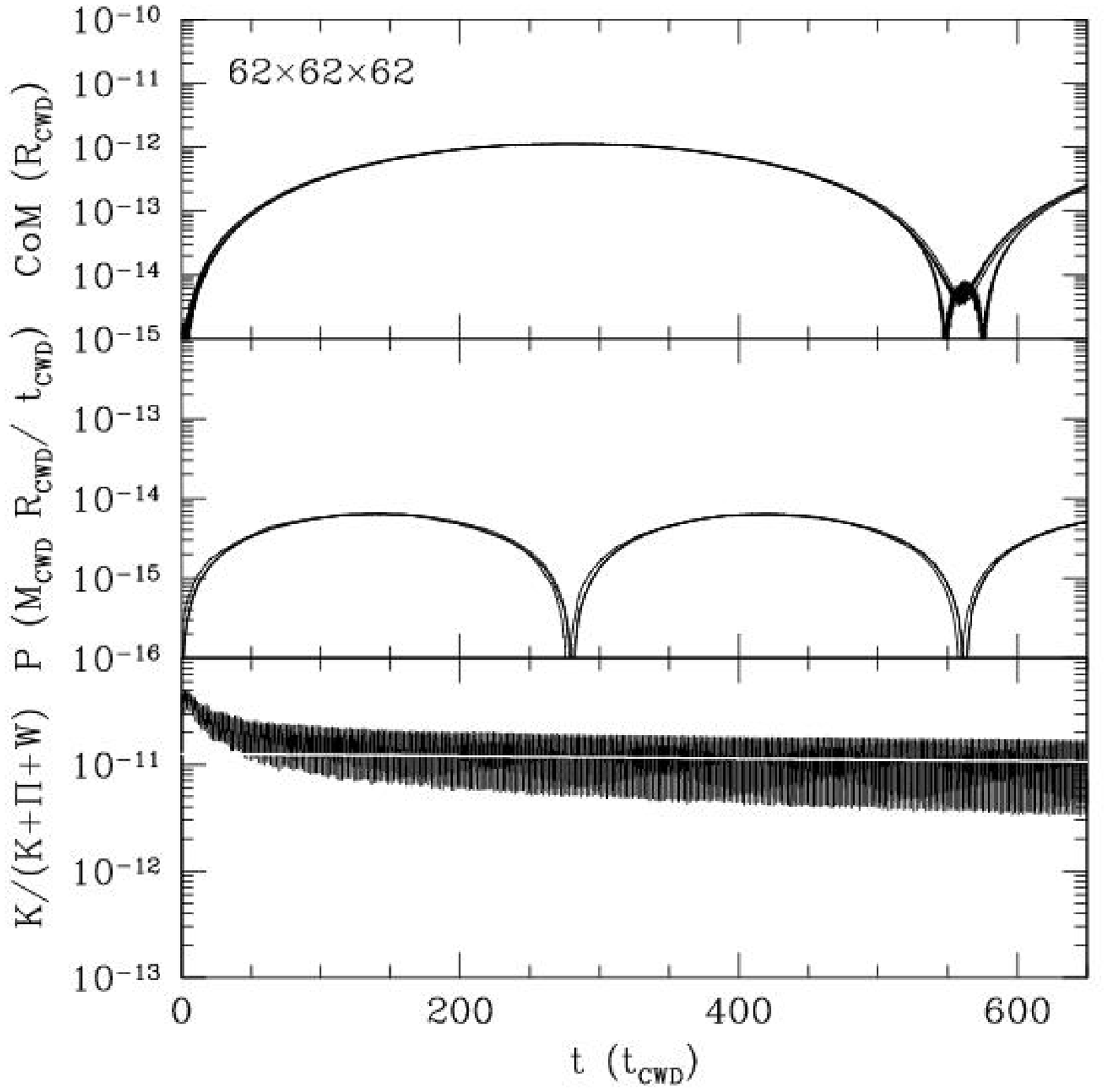}
\end{center}
\caption{Same as Figure \ref{sstars} for the case when a quadrupolar
  perturbation is present (note the difference in scales in comparison
  to that figure).  The white line drawn through the oscillations is
  for a $Q$ of approximately 6000.}
\label{qpstars}
\end{figure}
\begin{figure}
\begin{center}
\includegraphics[width=0.8\columnwidth]{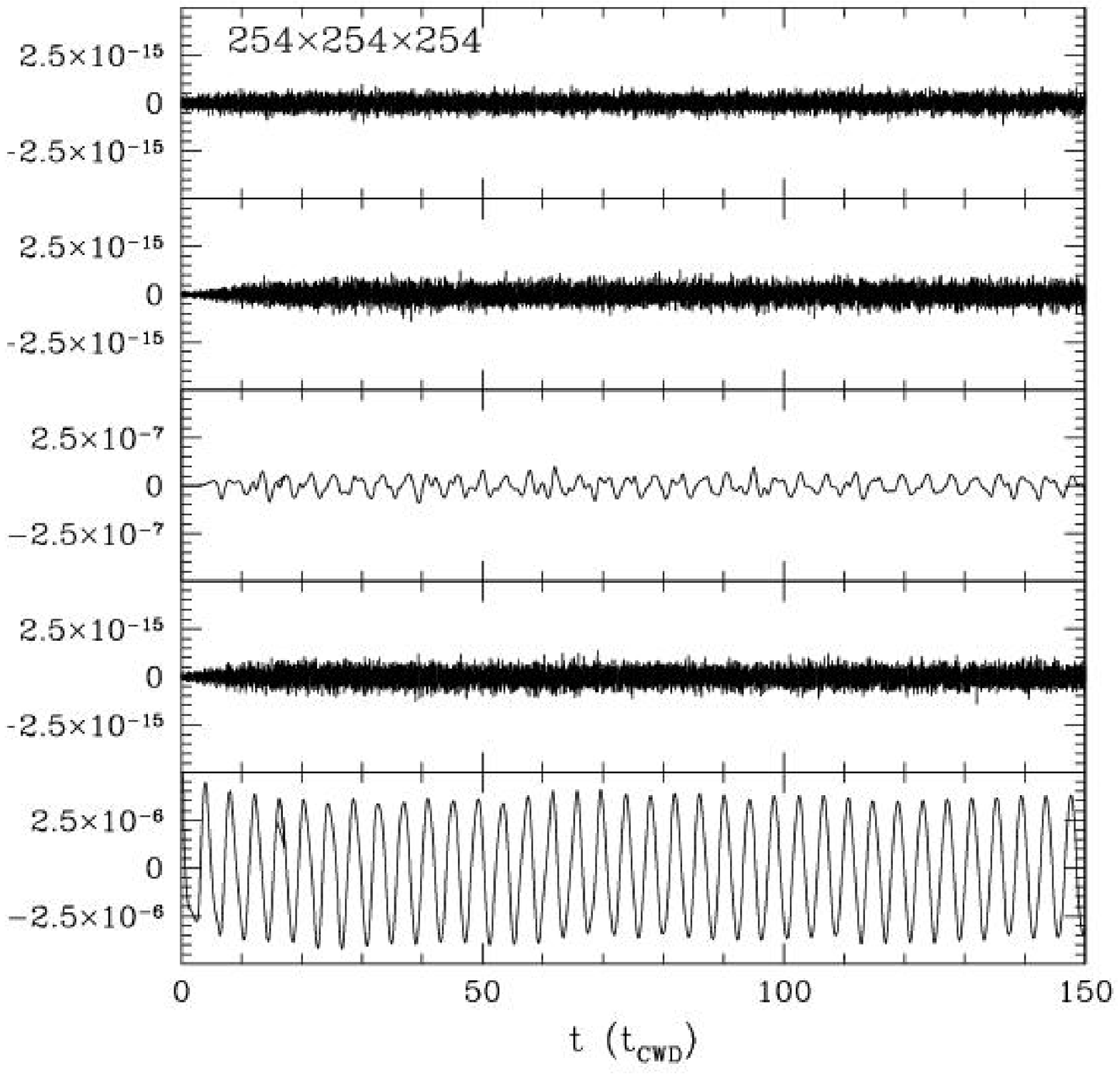}
\includegraphics[width=0.8\columnwidth]{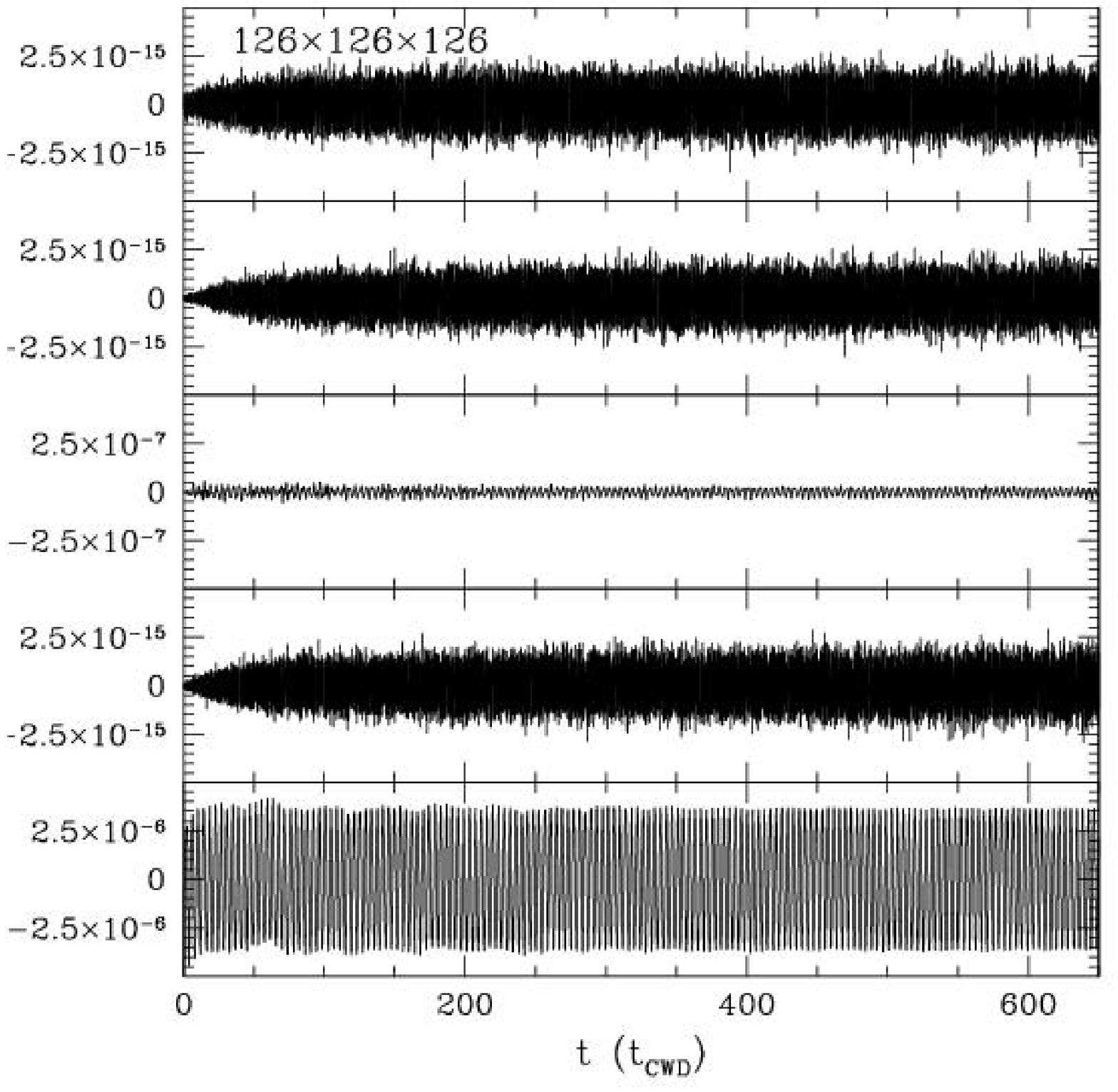}
\includegraphics[width=0.8\columnwidth]{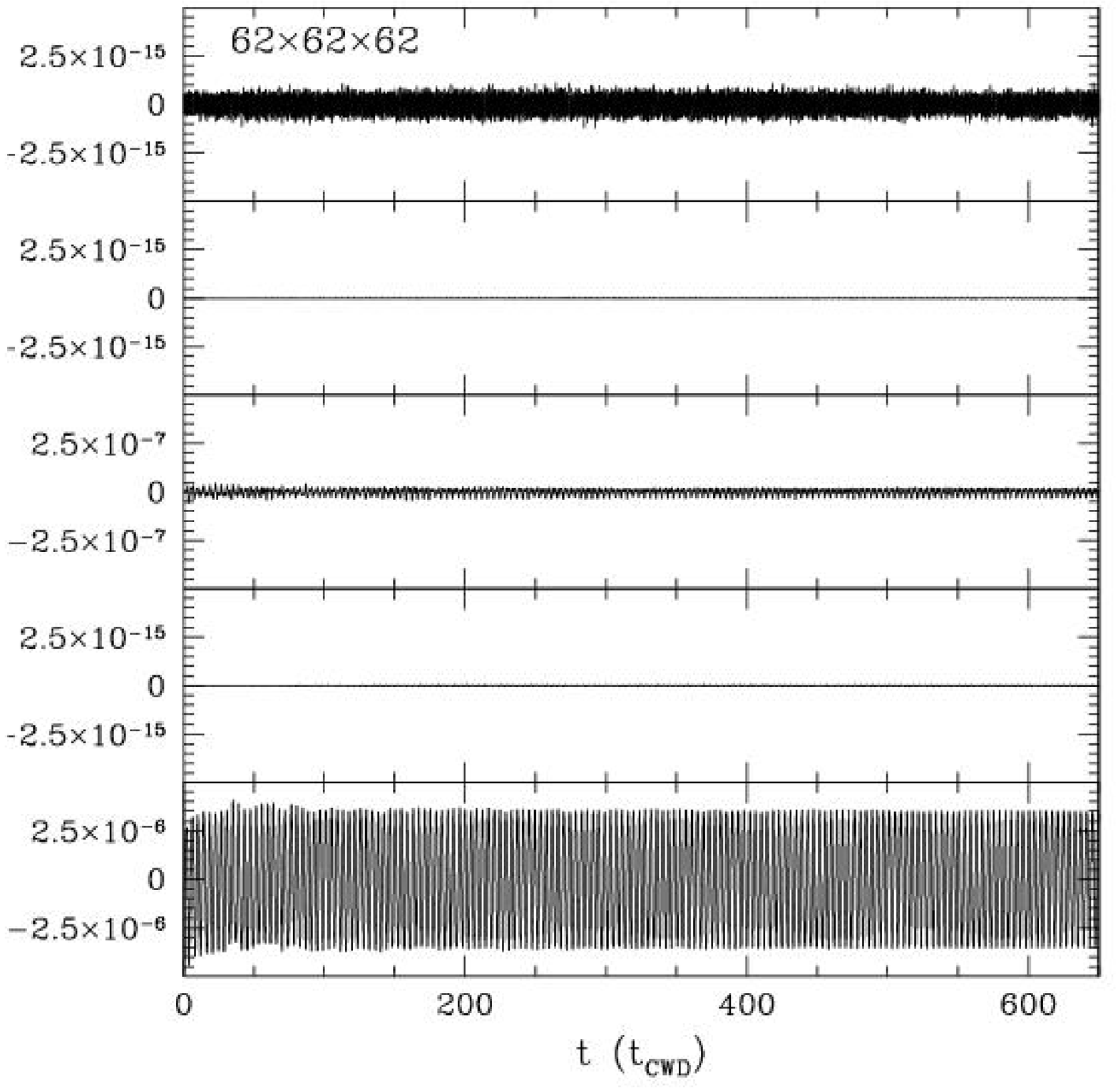}
\end{center}
\caption{The quadrupolar moments of the perturbed star for each of the
  resolutions considered in Figure \ref{qpstars}.  From top to bottom,
  the panels are the odd $m=2$, odd $m=1$, $m=0$, even $m=1$, and even
  $m=2$ moments.  Note the difference in scales of the different
  moments, namely that the even $m=2$ moment is two orders of
  magnitude larger than the $m=0$ moment (which is sourced by the
  Cartesian geometry) and nine orders of magnitudes larger than the
  others.
}
\label{qpmoments}
\end{figure}
We deformed the star by adding a fractional quadrupolar perturbation
to the density, \ie
\begin{equation}
\Delta\rho({\bf r}) = A\rho(r) Y^e_{22}(\theta,\phi)\,,
\end{equation}
where the amplitude, $A$, was chosen to be small ($10^{-4}$) so that
the resulting oscillation occurred in the linear regime.  This
initiated an even $m=2$ standing wave on the star.  Figures
\ref{qpstars} and \ref{qpmoments} show the resulting evolutions for
a number of grid resolutions.  The same diagnostics as those used to
demonstrate hydrostatic equilibrium are shown in Figure \ref{sstars}.
In this case as well, the centre-of-mass and momentum drift saturate
at levels well below those of interest.  Unlike hydrostatic
equilibrium, there now exists a non-vanishing kinetic energy.  It is
strongly harmonic and decays exponentially.  Because the initial
perturbation excited all of the even quadrupolar modes with $m=2$,
there are a number of distinct decay constants, with the slowest being
due to the $f2$ mode.  This exponential decay at late times may be fit
to estimate the numerical $Q$, found here to be on the order of $6000$.

\begin{figure}
\begin{center}
\includegraphics[width=\columnwidth]{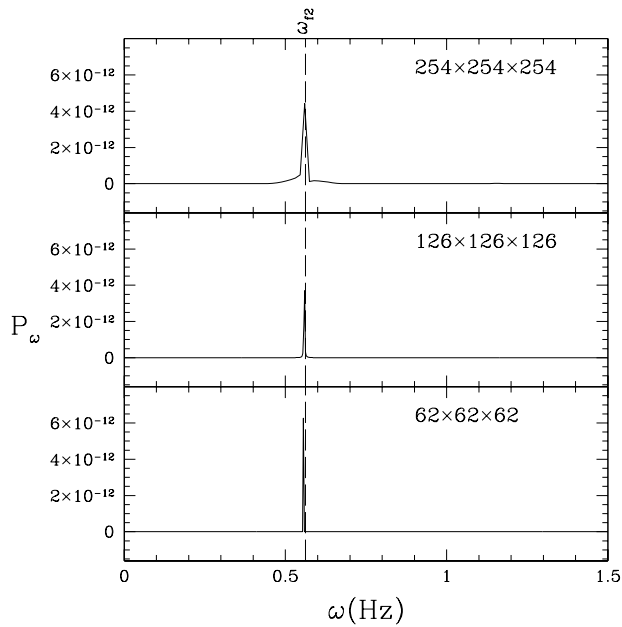}
\end{center}
\caption{Shown are the power spectra of the even $m=2$ quadrupolar
  moment as a function of angular frequency (using the mean squared
  amplitude normalisation).  As expected, for each grid resolution
  there is a strong spike coincident with the $f2$ mode frequency
  predicted for the HWD model.}
\label{qpfts}
\end{figure}
In Figure \ref{qpmoments}, the quadrupolar moments are shown.  The even
$m=2$ moment is strongly dominant as expected.  It also has a very
clear harmonic structure.  This may be Fourier analysed to produce the
dominant oscillation mode, as shown in Figure \ref{qpfts}.  In the
power spectrum of the even $m=2$ quadrupolar moment, there is a peak which
extends five orders of magnitude above the rest of the spectrum.  This
peak is clearly identifiable with the $f2$ mode, and appears to have
very nearly the frequency predicted by the HWD model.

\section{Conclusions} \label{C}
We have developed and tested a parallel, simple and fast hydrodynamics
code for multi-dimensional, self-gravitating, adiabatic flows.  Both
the advection terms and the solution for the self-gravity are greatly
simplified by the use of a uniform Cartesian geometry, ultimately
leading to explicit conservation of mass, entropy, and momentum to
nearly numerical accuracy.  The simplifying assumption of adiabaticity
and the absence of shocks eliminate the necessity for more
numerically expensive schemes, yielding an efficient code.

We have also applied our code to a number of standard diagnostic
problems in order to verify its physical correctness and limitations.
This has been done in a systematic fashion intended to test each aspect of
the code separately, including the advection scheme, the pressure
source terms, and finally the gravitational potential.  Finally we
have demonstrated the fitness of the code for the problem which motivated
its development: the study of tidally excited adiabatic oscillations
on white dwarfs.  This has been done in two stages: firstly, verifying the 
long term numerical stability of a white dwarf in hydrostatic
equilibrium, and, secondly, measuring the numerical quality factor
(found to be on the order of 6000) and the quadrupolar
fundamental mode frequency (found to be very nearly that predicted by
a linear mode analysis of the white dwarf model).  The details of tidally
exciting adiabatic oscillations with non-linear amplitudes on cold white dwarfs,
and their subsequent evolution will be discussed in a future publication.

\section*{Acknowledgements}
The authors would like to thank Anatoly Spitkovsky, Ruben Krasnopolsky, Michele
Vallisneri, Andrew MacFadyen, Joel Tohline, and Michael Norman for a
number of helpful conversations.  AEB would especially like to thank
Jim Stone for a number of useful suggestions regarding the
presentation of this material.  This work was supported under NASA
grant NAGWS-2837.

\bibliographystyle{mn2e.bst}\bibliography{hydro.bib}

\bsp

\end{document}